\documentstyle[amssymb,aps,psfig,epsf]{revtex}

\begin{document}
\title{Residual Resistance in a 2DES: A Phenomenological Approach.}
\author{A.F. Volkov$^{1,2}$ and V.V.Pavlovskii$^{2}$}
\address{$^{(1)}$Theoretische Physik III,\\
Ruhr-Universit\"{a}t Bochum, D-44780 Bochum, Germany\\
$^{(2)}$Institute of Radioengineering and Electronics of the Russian Academy%
\\
of Sciences, 103907 Moscow, Russia }
\maketitle

\begin{abstract}
We consider a simple phenomenological model of a semiconductor with absolute
negative conductance in a magnetic field. We find the form of the domains of
the electric field and current which arise as a result of an instability of
a uniform state. We show that in both Corbino disc and Hall bar samples the
residual conductance and resistance are negative and exponentially small;
they decrease exponentially with increasing length $L_{x,y}.$
\end{abstract}

\section{Introduction}

An interesting effect recently observed is that the resistance of a
microwave irradiated two dimensional system (2DES) in a magnetic field $B$
drops almost to zero in certain intervals of $B$ \cite{Klitzing,Zudov2}. In
magnetic fields too low for the observation of Shubnikov-de Haas
oscillations ($B\lesssim 0.1T$) the resistance $R_{xx}$ of a 2DES with
mobility $\simeq 3\times 10^{6}V/cm.\sec ^{2}$ begins to oscillate in the
presence of microwave irradiation \cite{Zudov1}. In samples with higher
mobility ($\simeq 2\times 10^{7}V/cm.\sec ^{2})$ the resistance oscillations
become more pronounced and the resistance drops almost to zero in some
intervals of $B$ \cite{Klitzing,Zudov2} lower than the fields at which\
Shubnikov-de Haas oscillations begin to appear. Sometimes the states of the
2DES with almost zero resistance are called the zero resistance state (ZRS).
These states have been observed in experiments on Hall bars \cite
{Zudov2,Klitzing}. Similar measurements have been recently carried out on
Corbino discs \cite{Zudov3}. In this case a zero conductance state rather
than the ZRS has been observed in the same intervals of the applied magnetic
field $B$. These experiments have drawn considerable attention, and during a
short period of time many papers appeared where possible mechanisms for the
observed effects were suggested \cite
{Phill,Yale,Shi,Andreev,AB,Raikh,Mikh,BHV,Dor,Mirlin,Ryzh,Aleiner}. The most
adequate mechanism seems to be the one based on the idea of absolute
negative conductance (ANC) \cite
{Yale,Shi,Andreev,AB,Raikh,BHV,Dor,Ryzh,Aleiner}. It was shown that a
microwave irradiated 2DES in a strong magnetic field ($\omega _{c}\tau >1,$
where $\omega _{c}$ is the cyclotron frequency, $\tau $ is the momentum
relaxation time) may have absolute negative conductance $\sigma _{xx}$. In
some papers the calculations were carried out for a small bias electric
field $E$ (a linear response). In Refs. \cite{Shi,Aleiner} the dependence $%
\sigma _{xx}(E)$ was discussed and the conclusion was made that the
dependence of $\sigma _{xx}(E)E_{x}$ on $E_{x}$ may have a N-shape form (see
Fig.1). The authors of Ref. \cite{Andreev} assumed that the function $%
E_{x}=\rho _{xx}(j_{x})j_{x}$ has the N-shape form and showed that the
states with negative differential conductance (NDC) are unstable with
respect to small perturbations. As a result of this instability, a
nonuniform distribution of the current density is assumed to arise and
current domains (or current filaments) should appear in the sample. In this
state the field $E_{x}$ is zero and a change in the total current leads for
example to an enlargement of the domain with positive current and to a
shrinking of the domain with negative current (the ZRS).

The possibility of a microwave induced ANC in semiconductors in a quantizing
magnetic field was suggested and discussed some decades \ ago \cite
{Ryzhii,Elesin,V'yurkov}. Around the same time phenomena in semiconductors
with N- and S-shape $I(V)$ characteristic were studied inensively both
experimentally and theoretically (see the review article \cite{VKUsp} and
the book \cite{bBonch} and references therein). In particiular, it was shown
that in semiconductors with the N-shape $I(V)$ characteristic generally
speaking moving domains of the electric field arise as a result of the
instability of a state with the NDC. In semiconductors with the S-shape $%
I(V) $ characteristics, domains (filaments) of the current density arise as
a result of the instability. In the presence of the electric or current
domains the $I(V)$ characteristic changes drastically. In the first case an
almost horizontal section appears on the $I(V)$ characteristics (zero
differential conductance), whereas an almost vertical section appears on the 
$I(V)$ curve in the second case (zero differential resistance). If a system
with S- or N-shape $I(V)$ characteristic is placed in a magnetic field, the
situation becomes more complicated: the form of the $I(V)$ curve now depends
on the measurement geometry (Hall bar or Corbino disc) \cite{BHV} and both
electric field and current domains may coexist. For example, if one assumes
the N-shape form for $j_{x}=\sigma _{xx}(E_{x})E_{x}$ dependence on $E_{x}$
in a Corbino disc ($E_{y}=0$), then in the Hall bar measurements with
increasing magnetic field the dependence $j_{x}(E_{x})$ acquires a
complicated form becoming S-shaped at $\omega _{c}\tau >1$ \cite{BHV}.

Strictly speaking, equations for macroscopic parameters (electric field,
electron density etc) corresponding to experimental samples \cite
{Klitzing,Zudov2,Zudov3} are not yet derived. In the recent paper \cite
{Aleiner} a kinetic equation for the distribution function is derived which,
in principle, could be applied to describe nonhomogeneous states. However
this task is very difficult. All previous publications on these
nonhomogeneous states are based on equations for macroscopic parameters.
Therefore, although the dependence of the ZRS or ZCS on the applied magnetic
field $B$ obtained theoretically \cite{Yale,Shi} qualitatively agrees with
experimental results \cite{Klitzing,Zudov2,Zudov3}, it is difficult to
convincingly conclude that the observed ZRS or ZCS are the result of
instability of an uniform state in the 2DES with the S- or N-shape $I(V)$
curve. Nevertheless even a phenomenological model of the S- or N-shape
characteristics allows one to make certain conclusions about the observed $%
I(V)$ curves. In this paper we admit a phenomenological model assuming a
N-shape $I(V)$ curve in the Corbino disc geometry and calculate the residual
resistance (Hall bar experiments) or conductance (Corbino disc experiments).
We show that in our model this resistance (conductance) depends
exponentially on the ratio of the width of the sample to the width of the
domain wall. Although our model does not correspond to the parameters of the
real samples used in experiments \cite{Klitzing,Zudov2,Zudov3}, it has a
physical meaning and allows us to make general statements about the
stability of the system, the form of electric and current domains and the
residual resistance of the system in a nonuniform state. In particular using
model one can answer the question whether the state observed in experiments
is dissipationless (as it happens in superconductivity and was anticipated
in Ref.\cite{Klitzing}) or not.

\section{Model}

As in Ref.\cite{BHV}, we assume that $\sigma _{xx}$ depends on the electric
field $E=\sqrt{E_{x}^{2}+E_{y}^{2}}$ so that in the absence of $E_{y}$ (the
Corbino disc) the dependence $j_{x}=\sigma _{xx}(E)E_{x}$ on $E_{x}$ has an
N-shape form (see Fig.1). Such a dependence was discussed in Ref.\cite{Shi}%
(see also \cite{Aleiner}). For example, this dependence may be obtained if
we take for $\sigma _{xx}(E)$ the expression

\begin{equation}
\sigma _{xx}(E)=\sigma _{\infty }-\frac{\sigma _{\infty }+\sigma _{0}}{%
1+(\sigma _{0}/\sigma _{\infty })(E/E_{0})^{2}}  \label{Sigma}
\end{equation}

{\large \bigskip }

where $\sigma _{\infty }$ is the conductance in strong fields ($E>>E_{0}$), $%
\sigma _{0}\equiv -\sigma _{xx}(0),$ and $\sigma _{xx}(E)$ equals zero at
the field $E_{0}.$ {\bf The electric field }$E_{0}${\bf \ as well as the
conductivity }$\sigma _{0}${\bf \ should be determined from a microscopic
theory (see Refs.\cite{Shi,Aleiner} and \cite{Ryzhii2}). For example,
according to Ref.\cite{Ryzhii2} on the order of magnitude }$E_{0}\gtrsim $%
{\bf \ \ }$20$ $V/cm$. In order to describe the form of domains, one has to
know a microscopic mechanism which determines the form of the $I(V)$ curve.
For example in Refs.\cite{VKUsp,VKZh} a superheating mechanism of the
S-shape current-voltage characteristic was considered. In this case the
width of the current domains is determined by the energy relaxation length.
We assume that in our model a characteristic length is determined by the
screening length (such a model was used to describe the Gunn effect \cite
{VKUsp,bBonch}). Therefore the current density can be written as

\begin{equation}
{\bf j}=\hat{\sigma}{\bf E}-e\hat{D}{\bf \nabla }n+(\epsilon /4\pi )\partial 
{\bf E}/\partial t  \label{Current}
\end{equation}

{\large \bigskip }

where $\hat{\sigma}$ is the conductivity tensor with diagonal components
equal to $\sigma _{xx}=$ $\sigma _{yy}\equiv \sigma $ and off-diagonal ones
equal to $\sigma _{xy}=-\sigma _{yx}\equiv \sigma _{H}$ , $\hat{D}$ is the
diffusion coefficient tensor which has a similar structure. The last term is
the displacement current. We assume that only $\sigma $ depends on the field 
$E$. The components $\sigma _{H}$ and $D,D_{H}$ are assumed to be
independent of $E$. {\bf At first glance the assumption about independence
of the diffusion coefficient on }$E${\bf \ violates the Einstein relation
between the mobility and diffusion constant. However this relation holds
only for systems in equilibrium. This is not the case for the system under
consideration. Of course the diffusion coefficient }$D${\bf \ depends on }$E$%
{\bf \ in some (unknown in our phenomenological appoach) way. This
dependence can be taken into account if we introduce a new ''potential'' in Eq.(6)
or a new electric field. Therefore a corresponding analysis for the case of the energy dependent
diffusion coefficient }$D(E)$ {\bf can be carried out in a similar way as
for the constant }$D.$ {\bf For simlicity we ignore this dependence because
it does not lead to qualitatively new results. Note that an analogous
problem was discussed in the theory of the Gunn effect (see the reviews } 
\cite{VKUsp,bBonch} {\bf and references therein). }

The electron concentration $n$ is related to the electric field via the
Poisson equation (instead of -e we write +e making a proper choice of the
electric field direction)

\begin{equation}
{\bf \nabla E}=(4\pi e/\epsilon )(n-n_{0})  \label{Poisson}
\end{equation}

We also assume that the thickness of the 2DES $d$ is larger (or of the
order) than the screening length $l_{scr}$. In this case in order to find a
relation between $n$ and ${\bf E}$, one has to solve Eq.(\ref{Poisson})
inside the sample. Otherwise one has to solve an equation for the electric
field (or potential) outside the sample taking into account corresponding
boundary conditions. In the case $l_{scr}\lesssim d$ Eqs.(\ref{Current},\ref
{Poisson}) yield correct results at least qualitatively. The main drawback
of our model is the assumption of a local relationship between the current
density ${\bf j}$ and electric field ${\bf E}$ which is true if the mean
free path $l$ is shorter than a characteristic length of the problem (in our
case the screening length). In the experiments one has the opposite
relationship between $l_{scr}$ and $l.$

\begin{figure}
\epsfysize= 5cm \vspace{0.2cm}
\centerline{\epsfbox{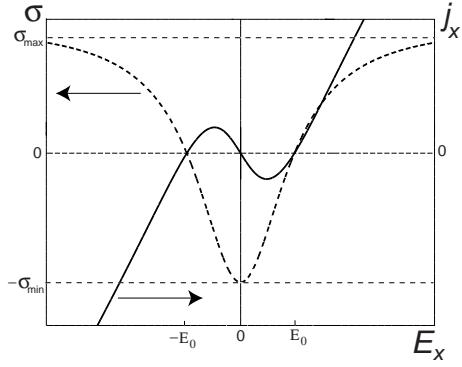}} \vspace{0.2cm}

\large{\caption{The assumed I(V) characteristic in the Corbino disc measurements (solid line) 
and the corresponding conductivity $\protect\sigma_{xx}$ as a function of $E_{x}$(dashed line).
Here $\protect\sigma_{min} = \sigma_0$ and $\protect\sigma_{max} = \sigma_\infty$.}}

\end{figure}

\section{The Corbino disc}

Let us consider first the simplest case of the Corbino disc geometry when $%
E_{y}=0$ (see Fig.2a).

\begin{figure}
\epsfysize= 5cm \vspace{0.2cm}
\centerline{\epsfbox{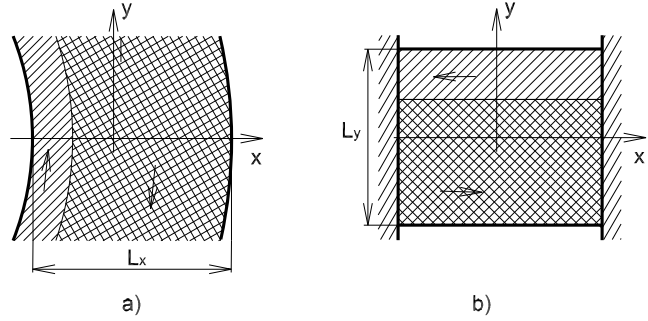}} \vspace{0.2cm}

\large{\caption{The Corbino disc (a) and Hall bar (b) samples (schematically).The region with diagonal stripes 
and the cross-hatch region mean domains with opposite directions of the current}}

\end{figure}

Eqs.(\ref{Current},\ref{Poisson}) can be rewritten in this case as follows

\begin{equation}
j_{x}=\sigma E_{x}-eD{\bf \partial }_{x}n+(\epsilon /4\pi )\partial
E_{x}/\partial t  \label{Current1}
\end{equation}

\begin{equation}
\partial _{x}E=(4\pi e/\epsilon )(n-n_{0})  \label{Poisson1}
\end{equation}

These equations practically coincide with equations used in the theory of
the Gunn effect \cite{VKUsp,bBonch}. Linearizing these equations with
respect to small fluctuations $E(x,t)=E_{x0}+\delta E_{x}(x,t)\ast \exp
(i\omega t-ikx)$, one can easily show that states with negative differential
conductance are unstable: $i\omega =-\partial j_{0}(E_{x})/\partial
E_{x}-k^{2}D,$ where $j_{0}=\sigma _{xx}(E_{x})E_{x}$ is the $I(V)$
characteristic in an uniform case. This dispersion relation shows that the
states with negative differential conductance are unstable with respect to
long-wave perturbations. The characteristic length of the problem (the width
of a domain wall) is determined by the relation: $l_{ch}=\sqrt{D/\mid \sigma
_{d}\mid }$, where $\sigma _{d}=\partial j_{0}(E_{x})/\partial E_{x}$ is the
differential conductance. The conclusion on the instability of the states
with a negative differential conductance was made a long time ago (see Refs 
\cite{ElesMan,Zakh,KogBB}). As a result of the instability, domains of the
electric field (in the x-direction) arise in the sample. In the y-direction
these domains may be regarded as the current density filaments: different
Hall currents flow in domains with opposite directions of the electric
fields. In order to find the form of the domains of the electric field $E$,
we exclude $n$ from Eqs.(\ref{Current},\ref{Poisson}) and arrive at the
equation

\begin{equation}
\frac{\partial ^{2}E_{N}}{\partial x_{N}^{2}}+\frac{\partial U}{\partial
E_{N}}-\gamma (\frac{\partial E_{N}}{\partial t_{N}}+j_{N0}(E_{N})\frac{%
\partial E_{N}}{\partial x_{N}})=0  \label{GenEq}
\end{equation}

Here $E_{N}=E_{x}/E_{0}$ is the normalised electric field, $x_{N}=x/l_{scr}$%
, $l_{scr}=\sqrt{D\epsilon /4\pi \sigma _{\infty }}$ is the characteristic
screening length, $\gamma =\epsilon E_{0}/4\pi en_{0}l_{scr},$ $%
j_{N0}(E_{N})=j_{0}(E_{N})/E_{0}\sigma _{\infty }$, $t_{N}=t/t_{ch}$, $%
t_{ch}=en_{0}l_{scr}/\sigma _{\infty }E_{0}$. The ''potential'' $U$ is
defined as

\begin{equation}
U(E_{N})=\int_{0}^{E_{N}}dE_{Nx}[j_{Nb}-j_{N0}(E_{Nx})]  \label{Potential}
\end{equation}

where $j_{Nb}=j_{b}/\sigma _{\infty }E_{0}$ is the normalised bias current, $%
j_{N0}(E_{N})=E\ \sigma (E_{N})/\sigma _{\infty }E_{0}$ is the normalised $%
I(V)$ curve (see Fig.1). The electron density $n_{0}$ in an uniform case is
the 3-dimensional electron concentration: it is related to the 2-dimensional
density $n_{20}$ via $n_{0}=n_{20}/d$, where $d$ is the thickness of the
2DES. We neglect the last term in Eq.(\ref{GenEq}) assuming that $\gamma $
is small. Let us estimate $\gamma $ for the samples used for example in the
experiments \cite{Klitzing,Zudov2,Zudov3}: for $n_{20}/d_{2}\approx 3\
10^{11}cm^{-2}/l_{scr}$, we get $\gamma =E_{0}/E_{ch}$, where $E_{ch}\approx
4\ 10^{4}V/cm$. Obviously the field $E_{0}$ is much smaller than the very
large value of $E_{ch}$. To find the form of stationary domains of the
electric field, we neglect the last term in Eq.(\ref{GenEq}). In the
following the analysis is similar to that presented in Ref. \cite{VKUsp}. We
integrate Eq.(\ref{GenEq}) and obtain the ''energy'' conservation law which
describes the phase trajectories

\begin{equation}
(1/2)E_{N}^{^{\prime }2}+U(E_{N})=U(E_{N1})  \label{Energy}
\end{equation}

The integration constant $U(E_{N1})$ determines the phase trajectory (see
Fig.3). Its choice depends on concrete boundary conditions. We assume that
at the boundaries $x=\pm L$ the spatial derivates $E_{N}^{\prime }\equiv
\partial E_{N}/\partial x_{N}$ is zero (the final result does not depend
qualitatively on particular boundary conditions). We are interested in a
solution of the form of two domains with opposite electric fields. The field 
$E_{N1}$ is connected with the length of the sample $L_{x}$ by the relation

\begin{equation}
L_{N}=\int_{-E_{N2}}^{E_{N1}}dE_{N}/E_{N}^{\prime }  \label{Length}
\end{equation}

{\large \bigskip }

where $L_{N}=L_{x}/l_{scr},L_{x}$ is the distance between the outer and
inner radious of the Corbino disc. The constant $E_{N2}$ is connected with $%
E_{N1}$ by the equation

\begin{equation}
U(-E_{N2})=U(E_{N1})  \label{E2}
\end{equation}

\begin{figure}
\epsfysize= 5cm \vspace{0.2cm}
\centerline{\epsfbox{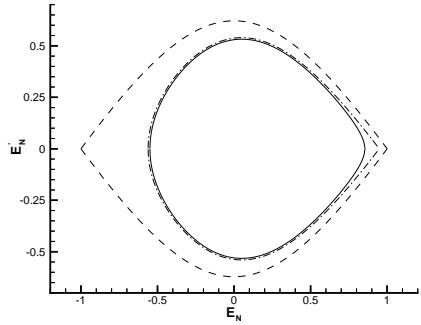}} \vspace{0.2cm}

\large{\caption{The phase trajectories. The solid
line shows a trajectory which describes the electric field distribution in
the considered finite sample in the presence of the bias current $j_{bN}$
(two domains with opposite directions of the fields). The maximum fields in
these domains are equal to $-E_{N2}$ and $E_{N1}$. The dash-dotted curve
shows a trajectory which corresponds to an infinite domain with a positive field 
and a finite domain with a negative field (soliton). Two infinte domains with opposite electric fields in the
absence of the bias current are described by the dashed trajectory.}}

\end{figure}

Our goal is to find a relation between the averaged field in the sample $%
E_{b}$ with the bias current $j_{b}.$ The relation $j_{b}(E_{b})$ is the
form of the current-voltage characteristic in the presence of domains. The
averaged field $E_{b}=\overline{E}_{N}E_{0}$ is given by

\begin{equation}
\overline{E}_{N}=L_{N}^{-1}\int_{-E_{N2}}^{E_{N1}}dE_{N}\
E_{N}/E_{N}^{\prime }  \label{AvE}
\end{equation}

The equations (\ref{Length},\ref{E2},\ref{AvE}) determines the $j_{b}(E_{b})$
characteristic, or the residual conductance in the presence of domains. The
function $E_{N}^{^{\prime }}(E_{N})$ in these equations is shown in Fig.3
(solid line). It is clear that if the characteristic width of the domain
wall (in our model $l_{scr}$) is much less than the length $L_{x}$, the
trajectory $E_{N}^{^{\prime }}(E_{N})$ should be close to the dash-dotted
trajectory (the soliton trajectory). We calculate the residual conductance $%
G_{res}$ assuming that $l_{scr}<<L_{x}$. In this case the fields $\pm
E_{N1,2}$ inside domains are close to $\pm 1$, the normalised bias current $%
j_{bN}$ is very small and in the main approximation the integrals in
equations (\ref{Length},\ref{E2},\ref{AvE}) can be calculated analytically.
Taking this into account, we find from Eqs.(\ref{Length},\ref{E2},\ref{AvE})

\begin{equation}
\delta E_{N2}^{2}\approx \delta E_{1}^{2}+4b  \label{Eq1}
\end{equation}

\begin{equation}
(\delta E_{N1}-b)(\delta E_{N2}+b)=\exp (-\sqrt{2a}L_{N})  \label{Eq2}
\end{equation}

\begin{equation}
\overline{E}_{N}=(1/L_{N}\sqrt{2a})\ln \frac{(\delta E_{N2}+b)}{(\delta
E_{N1}-b)}  \label{Eq3}
\end{equation}

Here $\delta E_{N1}=1-E_{N1}$, $\delta E_{N2}=-E_{N2}+1,$ $b=-j_{bN}/2a,$ $%
a=(1/2)\partial j_{N}(E_{N})/\partial E_{N}\rfloor _{E_{N}=1}=(1/2)\partial
\sigma _{N}(E_{N})/\partial E_{N}\rfloor _{E_{N}=1}. $ Note that in our
model the average field $\overline{E}_{N}$ is positive if the bias current $%
j_{bN}$ is negative, i.e. the resudial conductance is negative. We calculate
the residual conductance in the limit of small currents $j_{bN}<<1.$ If the
condition

\begin{equation}
\delta E_{N1}^{2}>>b  \label{Eq4}
\end{equation}

is satisfied, we get from Eqs.(\ref{Eq1},\ref{Eq2},\ref{Eq3})

$_{{}}$ 
\begin{equation}
\delta E_{N2}=\delta E_{N1}[1+2b/\delta E_{N1}^{2}]  \label{Eq1a}
\end{equation}

\begin{equation}
\delta E_{N1}=\exp (-\sqrt{2a}L_{N}/2)  \label{Eq2a}
\end{equation}

\begin{equation}
\overline{E}_{N}=2b\exp (\sqrt{2a}L_{N})/(\sqrt{2a}L_{N})  \label{Eq3a}
\end{equation}

Therefore the residual conductance has the form (we restore the dimensional
units)

\begin{equation}
G_{res}=-\sigma _{\infty }(a\sqrt{2a}L_{x}/l_{scr})\exp (-\sqrt{2a}
L_{x}/l_{scr})  \label{ResG}
\end{equation}

\bigskip

We see that $b$\ as well as $\delta E_{N1,2}$\ are exponentially small.
Therefore the assumed smallness of $\delta E_{N2}$\ is justified. The
formula (\ref{ResG}) is valid if the condition (\ref{Eq4}) is fulfilled.
This condition can be represented in the form

\begin{equation}
\overline{E}<<E_{0}(l_{scr}/L)\sqrt{2a}  \label{Eq4a}
\end{equation}

The formula (\ref{ResG}) shows that the residual conductance is small (we
assumed that $L_{x}>>l_{scr}$ ) and exponentially decreases with increasing
length $L_{x}$. Note that in a model more appropriate to the real
experimements \cite{Zudov2,Klitzing} the characteristic length might differ
from the screening length $l_{scr}$. If $\overline{E}$ exceeds the value of
the characteristic field in the rhs of Eq.(\ref{Eq4a}), the residual
conductance increases exponentially. Note that the obtained nonhomogeneous
state with a negative differential conductance is stable if the total
voltage is fixed (see Ref.\cite{VKUsp}).

\section{The Hall bar}

In this Section we consider the case of the Hall bar geometry (see Fig.2b).
In this case we have to write the equations for the currents $j_{x,y}$
having in mind that all quantities depend on $y$

\begin{equation}
j_{x}=\sigma E_{x}+\sigma _{H}E_{y}-eD_{H}{\bf \partial }_{y}n+(\epsilon
/4\pi )\partial E_{x}/\partial t  \label{CurrentX}
\end{equation}

\begin{equation}
j_{y}=\sigma E_{y}-\sigma _{H}E_{x}-eD{\bf \partial }_{y}n+(\epsilon /4\pi
)\partial E_{y}/\partial t  \label{CurrentY}
\end{equation}

Taking into account the Poisson equation (\ref{Poisson}) and the fact that $%
j_{y}=0,$ we obtain

\begin{equation}
j_{b}\approx \sigma _{H}E_{y}-(\epsilon D_{H}/4\pi ){\bf \partial }_{yy}E_{y}
\label{CurrentXa}
\end{equation}

\begin{equation}
j_{1}\approx \sigma (E_{y})E_{y}-(\epsilon D/4\pi ){\bf \partial }_{yy}E_{y}
\label{CurrentYa}
\end{equation}

Here $j_{b}=j_{x}(y)$ is the bias current density which depends on $y$, $%
j_{1}=\sigma _{H}E_{x}.$ The conductivity $\sigma (E)\approx \sigma (E_{y})$
because we assume that $\sigma _{H}>>\sigma $, and therefore $E_{y}>>E_{x}.$
As before we neglect small terms of the order $\gamma .$ The form of domains
is determined by Eq.(\ref{CurrentYa}) that coincides with Eq.(\ref{GenEq})
if the latter small term\ in Eq.(\ref{GenEq}) is neglected. We note the
electric field component $E_{x}$ does not depend on $y$ (this follows from
the Maxwell equation in the stationary state which we are interested in: $%
\nabla \times E=0$). The current $j_{b}$ is $y$-dependent; it has opposite
directions in different electric domains. Our aim is to find a relation
between the average current $\overline{j}_{b}$ and the electric field $E_{x}$%
, or the inverse relation $E_{x}(\overline{j_{b}}).$ From Eq.(\ref{CurrentXa}%
) we find

\begin{equation}
\overline{j}_{b}=\sigma _{H}\overline{E}_{y}  \label{CurrentXb}
\end{equation}

Calculating a relation between $j_{1}$ and $\overline{E}_{y}$ from Eq.(\ref
{CurrentYa}), as we did in the previous Section, we get

\begin{equation}
j_{1}=-G_{res}\overline{E}_{y}  \label{CurrentYb}
\end{equation}

Combining Eq.(\ref{CurrentXb}) and Eq.(\ref{CurrentYb}), we finally find

\begin{equation}
E_{x}=-(G_{res}/\sigma _{H}^{2})\overline{j}_{b}  \label{I-VHbar}
\end{equation}

This result means that the residual resistance $R_{res}$ is equal to $%
R_{res}=(G_{res}/\sigma _{H}^{2}),$ i.e., the residual resistance also is
exponentially small in the Hall bar samples measurements. Note that in the
expression for $G_{res}$ ({\large \ref{ResG}}) {\large \ }the length $L_{x}$%
\ should be replaced by $L_{y}.$ The $I(V)$ characteristics for a 2DES with
two current (electric field) domains is shown in Fig.3 of Ref.\cite{BHV}.

\section{Conclusions}

Using a simple phenomenological model, we have calculated the residual
conductance (resistance) of a 2DES. We have shown that in measurements on
both the Corbino disc and the Hall bar the residual conductance and
resistance are exponentially small; they decrease exponentially with
increasing length $L_{x,y}.$ In our model the residual conductance and
resistance are negative. A direct comparision of our results to experimental
data meets difficulties because, as we mentioned before, parameters of the
model do not correspond to real samples. Nevertheless one can note that the
observation of a small negative resistance was observed in Ref.\cite{Willett}%
. It would be of interest to measure the dependence of $R_{res}$ (in Hall
bars) or $G_{res}$ (in Corbino discs) on the length $L_{y}$ or $L_{x}$. This
can shed light on the applicability of a model based on the negative
conductance to the phenomenon observed in the experiments \cite
{Klitzing,Zudov2,Zudov3}.

{\bf Note that we considered a limiting case }$\sigma _{H}>>\sigma ${\bf .
For an arbitrary ratio }$\sigma _{H}/\sigma ${\bf \ a linear analysis of the
superheating instability in semiconductors with the S- or N shape I(V)
characteristics has been carried out by Kogan \cite{Kogan}. It was shown
that, as it happens in plasma \cite{Dykhne}, the superheating instability
results in the appearance of oblique current filaments (electric field
domains). Only in the limit }$\sigma _{H}/\sigma >>1${\bf \ these filaments
are parallel to the bias current. Analysis of oblique domains is much more
difficult problem as one needs to solve nonlinear equations in two
dimensions with corresponding boundary conditions. It is clear however that
if the voltage difference }$V${\bf \ is measured between peripheral
contacts, the Hall voltage }$V_{H}${\bf \ also contributes to }$V${\bf \ in
the case of oblique domains. Since }$V_{H}${\bf \ changes sign with changing
the magnetic field direction, the voltage }$V${\bf \ also can change sign.
Therefore the contribution of the Hall voltage }$V_{H}${\bf \ to }$V${\bf \
might be a reason of the sign change of }$V${\bf \ with changing the
magnetic field direction observed in recent experiments \cite{Willett}.}

{\bf Note added}: After the preparation of this manuscript \cite{VP} we
became aware of Ref.\cite{Merz} in which similar ideas were elaborated.

\end{document}